\documentclass[letterpaper, 10 pt, conference]{ieeeconf} 
\usepackage{amsmath}    
\usepackage{amsfonts}
\usepackage{cancel}
\usepackage{algorithm}
\usepackage{algorithmic}
\usepackage{empheq}
\usepackage{float}
\usepackage{booktabs}
\usepackage{siunitx}
\usepackage{multirow}
\IEEEoverridecommandlockouts                              % This command is only needed if  % you want to use the \thanks command
\usepackage{graphicx}
\usepackage{cite}      % nice compressed numeric cites like [3]–[5]
\usepackage{xcolor}
\title{\LARGE \bf
Long-Term Mapping of the Douro River Plume \\with Multi-Agent Reinforcement Learning
}

\author{Nicol\`o Dal Fabbro$^{1}$, Milad Mesbahi$^{1}$, Renato Mendes$^{2,3}$, Jo\~{a}o Borges de Sousa$^{2,3}$, George J. Pappas$^{1}$ 
\thanks{N. Dal Fabbro acknowledges the support from his AI x Science Postdoctoral Fellowship awarded through the University of Pennsylvania's IDEAS, DDDI, and Penn AI initiatives. R. Mendes and J. Borges de Sousa acknowledge the support of Fundação para a Ciência e a Tecnologia (FCT) for its financial support to LAETA via the project UID/50022/2025.} %<-this  stops a space
\thanks{$^{1}$Department of Electrical and Systems Engineering, University of Pennsylvania, USA. Correspondence to:  {\tt\small ndf96@seas.upenn.edu}.}%
\thanks{$^{2}$Laboratório de Sistemas e Tecnologia Subaquática (LSTS), Faculdade de Engenharia da Universidade do
Porto, Portugal}%
\thanks{$^{3}$ Laboratório Associado de Energia, Transportes e Aeronáutica (LAETA), INEGI, Porto, Portugal}
}

\begin{document}

\maketitle
\thispagestyle{empty}
\pagestyle{empty}

%%%%%%%%%%%%%%%%%%%%%%%%%%%%%%%%%%%%%%%%%%%%%%%%%%%%%%%%%%%%%%%%%%%%%%%%%%%%%%%%
\begin{abstract}
We study the problem of long-term (multiple days) mapping of a river plume using multiple autonomous underwater vehicles (AUVs), focusing on the Douro river representative use-case. We propose an energy - and communication - efficient multi-agent reinforcement learning approach  
in which a central coordinator intermittently communicates with the AUVs, collecting measurements and issuing commands. Our approach integrates spatiotemporal Gaussian process regression (GPR) with a multi-head Q-network controller that regulates direction and speed for each AUV. Simulations using the Delft3D ocean model demonstrate that our method consistently outperforms both single- and multi-agent benchmarks, with scaling the number of agents both improving mean squared error (MSE) and operational endurance. In some instances, our algorithm demonstrates that doubling the number of AUVs can more than double endurance while maintaining or improving accuracy, underscoring the benefits of multi-agent coordination. Our learned policies generalize across unseen seasonal regimes over different months and years, demonstrating promise for future developments of data-driven long-term monitoring of dynamic plume environments. 
\end{abstract}

%%%%%%%%%%%%%%%%%%%%%%%%%%%%%%%%%%%%%%%%%%%%%%%%%%%%%%%%%%%%%%%%%%%%%%%%%%%%%%%%
\section{Introduction}
Monitoring dynamic coastal environments in real-time is a persistent challenge in both environmental science and robotics \cite{raine2025, pastra2025}. Coastal waters evolve under the interplay of currents, winds, tides, and river discharge, producing transient patterns that are difficult to capture with conventional methods. A prominent example of this dynamism is a river plume: a buoyant outflow of freshwater that extends into the ocean.
River plumes, which are defined by their steep salinity gradients, are integral to the mixing processes that impact fisheries, water quality, and the spread of pollutants in the coastal areas \cite{mendes2014observation, horner2015mixing}. However, given their large extension in the ocean (hundreds of square kilometers) and rapid variability, 
tracking and mapping river plumes via fixed sensors or manned surveys is usually impractical \cite{Teixeria2021}.

Autonomous underwater vehicles (AUVs) offer a promising alternative. Indeed, AUVs can adapt their trajectories to the evolving conditions of aquatic environments, collect measurements across large spatial domains, and operate continuously over extended periods. Yet using AUVs for long-term plume monitoring introduces several challenges: (i)~the salinity field that characterizes the plume, a scalar function of space and time, evolves on the same timescale as vehicle motion, meaning that the process shifts significantly while measurements are being collected; (ii)~ocean currents in the plume area can dramatically hinder the AUVs mobility; (iii)~endurance is constrained by onboard energy reserves, resulting in a trade-off between coverage and longevity; (iv)~inter-agent communication in the ocean plume waters is severely constrained due to horizontal and vertical density variations, so AUVs usually need to dwell at the sea surface to transmit information\cite{yan2023formation}. Hence, coordinating multiple AUVs to map a river plume presents unique challenges, which we aim to address in this work.

The Douro River plume, where freshwater from the river with the greatest discharge in Portugal’s northwest coast enters the Atlantic, exemplifies these challenges and motivates this work. Its shape, extent, and orientation vary widely with river discharge, wind forcing, and tidal cycles. 
During high-flow periods, 
strong freshwater outflow drives the plume offshore, where it can be advected for tens of kilometers along the coast \cite{vieira2000}. 
More than 500,000 residents of Porto and Vila Nova de Gaia live along the Douro banks and draw economic value from the estuary. Hence, timely and accurate salinity maps, which capture the state of the plume, are highly valuable for informed management \cite{estevez2019, friedland2025characterization}.
\begin{figure}
    \centering
    \includegraphics[width=0.7\columnwidth, trim ={0.4cm 0.4cm 0.4cm 0.4cm}, clip]{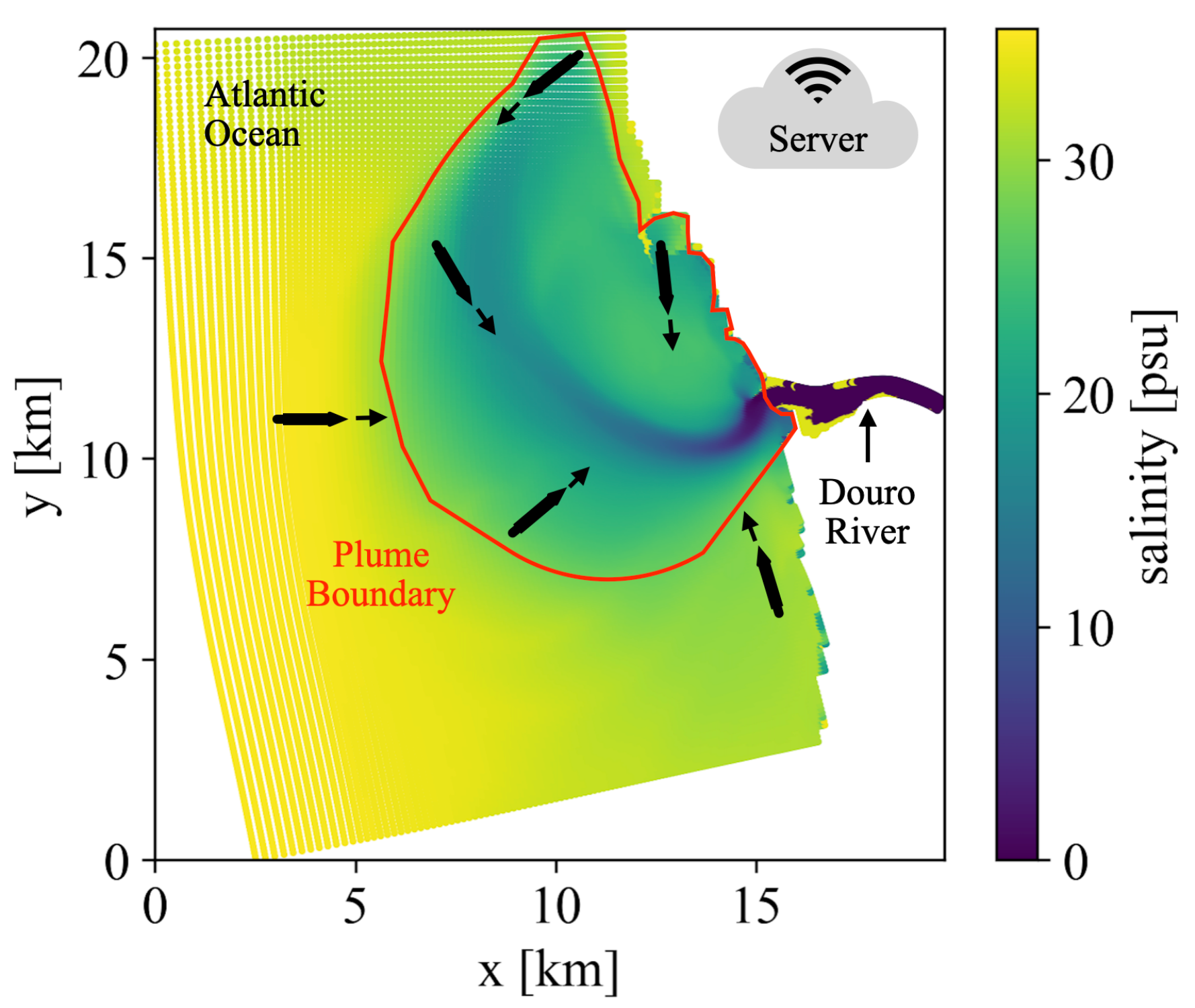}
    \label{fig:placeholder}
    \caption{Plume monitoring setting. The Douro River discharges into the Atlantic Ocean, generating a dynamic salinity plume (whose edge is represented by a red boundary). Multiple AUVs (black arrows) collect trajectory-constrained measurements and intermittently communicate with a central server to coordinate.}
\end{figure}
%\vspace{pt}

\begin{figure*}[t]
\centering
\includegraphics[width=0.95\textwidth]{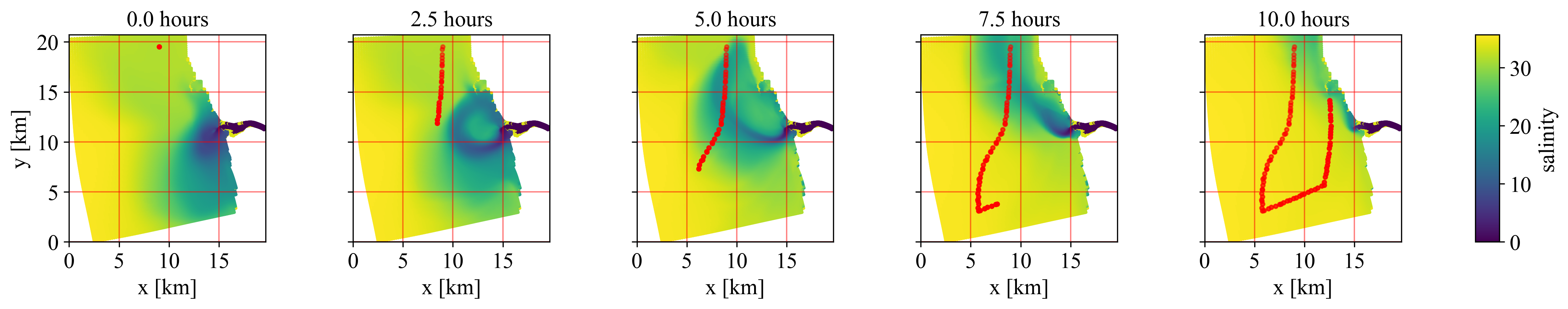} % full page width
\caption{Illustrative example of 10 hours of spatiotemporal evolution of the Douro river plume and AUV mobility (red trajectory), during March 2, 2018. The AUV uses the propulsion that provides the nominal speed of $1 m/s$ (in absence of ocean flow). Note that the AUV mobility is impacted by the currents.
}
\label{fig:frames_seq}
\end{figure*}
\textbf{Contributions.} In this paper, we propose and evaluate a cooperative data-driven multi-agent control %reinforcement learning~(RL) 
framework for long-duration (multiple days), energy-aware mapping of the Douro river plume. In our proposed system architecture, a server remotely coordinates a fleet of multiple light AUVs (LAUVs)~\cite{Teixeria2021}, handling computationally heavy decisions and requiring only minimal, intermittent radio communication with the vehicles, which devote energy to low-level navigation and sensing.
Within this setting, we propose an algorithm that combines a simple and reliable non-parametric Gaussian process regression (GPR) estimator with a reinforcement learning (RL) decision-making module based on a deep Q-network (DQN). 
Since most of the energy consumption of the considered LAUVs comes from the vehicle propulsion, we introduce a multi-head DQN architecture that decouples per-agent direction and speed decisions, paired with a reward function that allows us to control the tradeoff between estimation accuracy and energy efficiency.

We evaluate our approach and conduct a simulation study using a numerical model implementation of the Douro estuary with the open-source Delft3D ocean model~\cite{gerritsen2007validation, sousa2018}, which combines real coastal geometry and historical environmental data of the study area to generate high fidelity, realistic salinity and flow fields under distinct wind, tidal, and discharge regimes \cite{sousa2018}. Our results show that the proposed approach (i) effectively maps the plume and generalizes to unseen conditions across different months and years; (ii) consistently outperforms baselines and benchmarks in both single- and multi-agent settings; (iii) effectively leverages multi-agent collaboration, both in terms of estimation accuracy and fleet endurance, for example scaling from 3 to 6 vehicles more than doubling mission lifetime.

\textbf{Related work.}
Early coastal-robotics surveys rely on pre-planned coverage (lawn-mower transects, yo-yo depth cycling) \cite{Zhang2013}, approaches still common operationally \cite{Hwang2019} but prone to oversampling and poor adaptability. %A major advance cast the task as an 
Classic adaptive approaches frame the problem as informative path-planning (IPP) with Gaussian process surrogates \cite{Das2013}, selecting waypoints that maximize posterior variance reduction or mutual information under travel constraints. Longer horizons have been pursued via rapidly exploring random trees embeddings \cite{Hollinger2014} and non-myopic combinatorial methods such as GP-aware MILP and submodular branch-and-bound \cite{Dutta2022}, yet these approaches generally rely on fixed graph discretizations, offer limited online re-planning, and scale poorly. Within plume monitoring, 
recent works focus on classification-oriented objectives such as expected integrated Bernoulli variance (EIBV) minimization, which prioritize sampling along uncertain river-ocean interface zones \cite{Ge2023, Berild2024}. While relevant, these methods are single-vehicle, single-step sighted, energy-unaware, and limited to short missions ($\sim$4h).
Multi-AUV studies, on the other hand,
remain largely heuristic and oriented toward short-term front following rather than long-term full mapping \cite{Teixeria2021}.
Model-free RL offers an alternative to traditional IPP and plume monitoring methods. Convolutional neural network (CNN)-based DQN and multi-head DQN surpassed lawn-mower baselines for multi-agent lake monitoring \cite{Luis2021}, though only on static processes and coarse grids.
%with Brownian idleness dynamics and distance-only motion costs that ignore currents and energy.
Similarly, the work in~\cite{zhao2023adaptive} also explored the idea of combining GPR and RL on static, synthetic datasets.
Aerial domains show similar patterns in static pollution plumes, wildfire tracking, 
and radio maps~\cite{Assenine2023, Krij2023, Viseras2021}.

\section{Problem Formulation}
We model the salinity field of the Douro river plume as a spatiotemporal scalar function 
\begin{equation}
    f:\Omega \times \mathbb{R}_+ \to \mathbb{R}
\end{equation}
where $\Omega \subset\mathbb{R}^2$ denotes a planar domain and $\mathbb{R}_+$ is time. In this work, we consider 2-dimensional mapping of the plume, which is usually considered the region in $\Omega$ where salinity is lower than the open-ocean background level $f_{ocn}$ ($\approx35$ psu). For a chosen threshold $\zeta > 0$, we can define the plume at time $t$ as
\begin{equation}\label{eq:plume_def}
  \mathcal{P}(t) = \{\, x \in \Omega : f_{\mathrm{ocn}} - f(x,t) \ge \zeta \,\}.  
\end{equation}
Such $\mathcal{P}(t)$ evolves dynamically and, depending on environmental conditions (such as wind, ocean flow, discharge level, and tidal cycle), at a speed comparable to AUVs mobility ($1$m/s), as we illustrate in Fig.~\ref{fig:frames_seq}.

Our objective in this work is to construct estimates $\hat{f}$ of the salinity field $f$ over time, based on measurements collected by a fleet of $N$ AUVs deployed over~$\Omega$. A vehicle $n$ follows the following mobility model~\cite{aguiar2018}: %driven by its commanded propulsion and the ocean flow field
\begin{equation}\label{eq:mobility_model}
    \dot{x}^{(n)}(t) \;=\; u^{(n)}(t) + c\!\left(x^{(n)}(t),\,t\right),
\end{equation}
where $x^{(n)}(t)\in\Omega$ is the AUV position, $u^{(n)}(t)\in\mathbb{R}^2$ is its commanded velocity, and $c(x, t)$ is the ocean flow, described by a time-varying vector field
\begin{equation}
    c:\Omega \times \mathbb{R}_+ \to \mathbb{R}^2, \quad c(x,t) = (c_{\text{lon}}(x,t), c_{\text{lat}}(x,t)),
    \label{ocean-cur}
\end{equation}
\begin{figure}[t]
\centering
\includegraphics[width=\columnwidth, trim ={1cm 0.2cm 2cm 0cm}, clip]{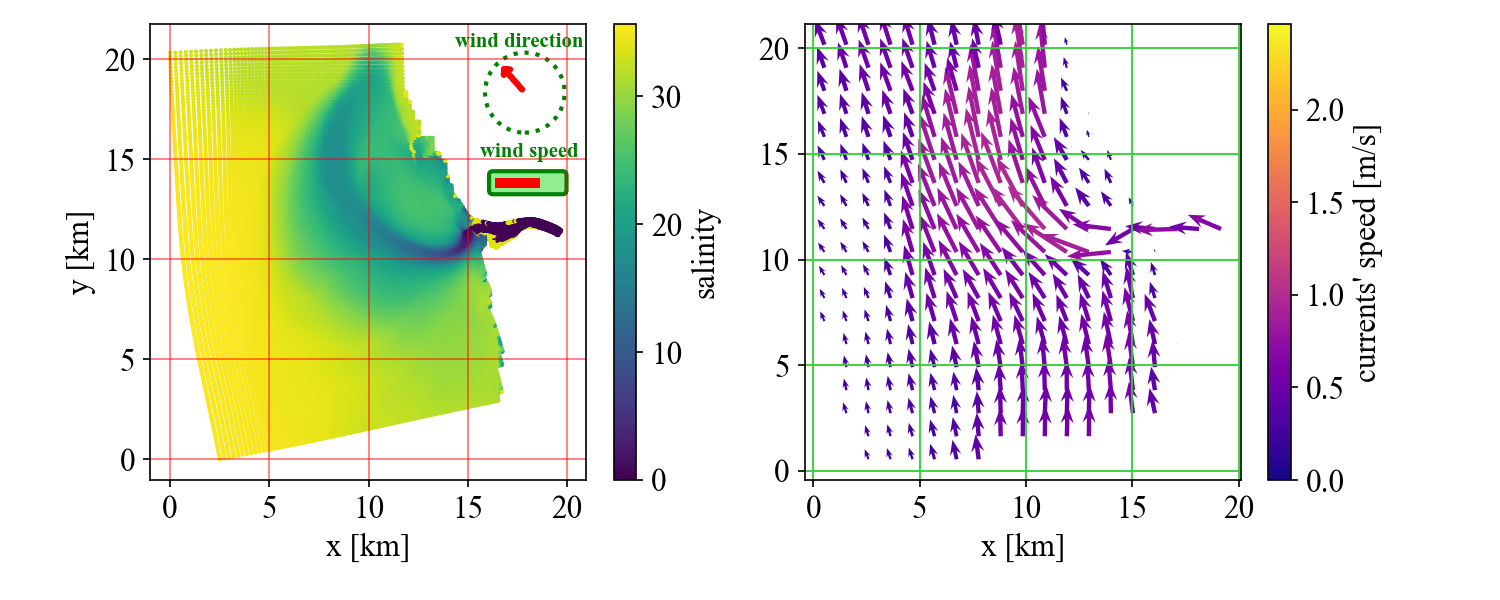} 
\caption{Salinity map of the Douro river plume (on the left) and a visualization of the ocean flow (on the right). Note the correlation between the salinity map and the currents' distribution, and that the speed of the currents gets above $1m/s$, namely the nominal speed of the AUVs.}
\label{fig:flow}
\end{figure}
where $c_{\text{lon}}(x,t)$ and $c_{\text{lat}}(x,t)$ are the longitude and latitude velocity components of the current, respectively.

To construct the estimate $\hat{f}$ of the unknown function $f$ over space and time, we want to collect measurements of the function $f$ with the AUVs. Given that the AUVs move according to~\eqref{eq:mobility_model}, the measurements that we can use for the estimation are \textit{constrained} in space and time by their $N$ trajectories. To formally state our problem formulation, we need to introduce some notation. Let us first define a vector of discrete time slots $\sigma = [0, \Delta, 2\Delta, ..., T\Delta]$ which we index with $\mathcal{K} = \{0, 1, ..., T\}$. In the rest of the paper, we are interested in updating the estimate $\hat{f}$ over the slots $\sigma[1], ..., \sigma[T]$ with a slot granularity of $\Delta = 30$ minutes. 

Let $\gamma^{(n)}_k:[(k-1)\Delta, k\Delta]\rightarrow\Omega$ denote the trajectory navigated by agent $n$ during time slot $k$ and let $\Gamma^{(n)}_k = \left\{\left(\gamma^{(n)}_k(t), t\right): t\in[(k-1)\Delta, k\Delta]\right\}$. 
% $\Gamma(A, B)$ denote the set of admissible curves connecting points $A$ and $B$. At time slot $k \in\mathcal{K}$, agent $n$ has navigated 
% %from location $x^{(n)}((k-1)\Delta)$ to $x^{(n)}(k\Delta)$ 
% along trajectory $\gamma_k^{(n)}\in\Gamma(x^{(n)}((k-1)\Delta), x^{(n)}(k\Delta))\subset\Omega$. %\footnote{$\Gamma(A, B)$ denotes the set of admissible curves connecting $A$ and $B$}.
We denote the sampled set of space-time coordinates along $\Gamma_k^{(n)}$ by 
\begin{equation*}
Z_k^{(n)} = \left\{\left(x^{(n)}\left(t_{k, 1}^{(n)}\right), t_{k, 1}^{(n)}\right), ..., \left(x^{(n)}\left(t_{k, z_{k}^n}^{(n)}\right), t_{k, z_{k}^n}^{(n)} \right)\right\},
\label{measurement}
\end{equation*}
with $z_{k}^n = |Z_{k}^{(n)}|$. Note that $(k-1)\Delta = t_{k, 1}^{(n)} \leq \cdots \leq  t_{k, z_{k}^n}^{(n)} = k\Delta$. % and that $Z_k^{(n)}\subset\gamma_k^{(n)}$, with $x((k-1)\Delta) = x_{k, 1}^{(n)}$, $x(k\Delta) = x_{k, z_{k,n}}^{(n)}$.
Let the set of space-time locations visited up to time slot $k$ by agent $n$ be denoted by $D_k^{(n)} = \{Z_{1}^{(n)}, ..., Z_{k}^{(n)}\}$ and let $D_k^{N} = \{D_k^{(n)}\}_{n = 1}^N$ be the overall set of coordinates visited by the fleet up to time $k$. 
%Note that $D_k^{(n)}$ belongs to the long-term trajectory of agent $n$, which is the concatenation of $\{\gamma_l^{(n)}\}_{l=1}^{k}$.
%$\Gamma_k^{(n)} = \{\gamma_l^{(n)}\}_{l = 1}^k$. 
We denote the noisy fleet-measured salinity values up to slot $k$ by 
\begin{equation}
f_k^N = \{f(x, t)+\epsilon_{x, t} :(x, t)\in D_k^N\},
\end{equation}
where $\epsilon_{x,t}$ is a generic noise term. Given space-time coordinates $A$, let $f(A) =\{f(x, t)+\epsilon_{x, t} :(x, t)\in A\}$ be the corresponding measurements, 
%and we denote the overall data collected up to time slot $k$ by 
and let $\mathcal{M}_k^N = \{D_k^N, f_k^N\}$.

Given a measurement-based predictor $\hat{f}(\cdot|\mathcal{M}_k)$ for the salinity field $f(\cdot)$, and an evaluation grid $G\subset\Omega$, we formulate the long-term mapping problem as follows:
\begin{empheq}[left=\parbox{1.8cm}{\centering\bfseries Long-Term\\Mapping\\Problem}\empheqlbrace, right=,]{equation}
\label{eq:setting-obj}
\begin{aligned}
\min_{\{\mathcal{M}_k^N\}_{k=1}^T}\ \  
& \frac{1}{|G|\,T}\sum_{k=1}^{T}\sum_{x\in G}e_k^2(x) \\[0.25em]
\text{s.t.}\ \ 
& Z_k^{(n)}\subset \Gamma_k^{(n)} \subset \Omega, \quad \forall k, n,
\end{aligned}
\end{empheq}
where note that in the case of the AUVs sampling the constraints are satisfied by construction, and
\begin{equation}
    e_k(x) = f\left(x,\sigma[k]\right)-\hat{f}\left(x,\sigma[k]\mid \mathcal{M}_k^N\right),
\end{equation}
recalling that $\mathcal{M}_k^N = \{D_k^N, f_k^N\}$, and $D_k^N = \{D_k^{(n)}\}_{n=1}^N = \{\{Z_l^{(n)}\}_{l=1}^k\}_{n=1}^N$.
\vspace{0.1cm}\\\indent
\textbf{Discussion and challenges. } Given that the function $f(\cdot)$ is unknown and measurements become available only in real time, problem~\eqref{eq:setting-obj} must be solved online, sequentially building the set $\mathcal{M}_k^N$, for $k = 1, ..., T$. The first main challenge in doing this is due to the physical constraints in building the trajectories $D_k^N$, \textit{combined with the fact that the field $f(\cdot, t)$ changes in time at a speed comparable to the speed with which AUVs can move in the water}. Mathematically, the new locations $\{Z_k^{(n)}\}$ that we can add to the set $D_{k-1}^N$ can only cover a limited space which is comparable to the space that the plume $\mathcal{P}(t)$ in~\eqref{eq:plume_def} covers in the same amount of time. This is also referred to as the \textit{lack of synopticity} in the trajectory-constrained measurements~\cite{gomis2005errors}. We illustrate this in the sequence of frames in Fig.~\ref{fig:frames_seq}, where we plot the evolution of the field $f(\cdot, t)$ over 10 hours and the corresponding mobility of one AUV across the area of interest.  
\begin{figure}[t]
\centering
\includegraphics[width=0.65\columnwidth, trim ={0cm 0.6cm 0cm 0cm}, clip]{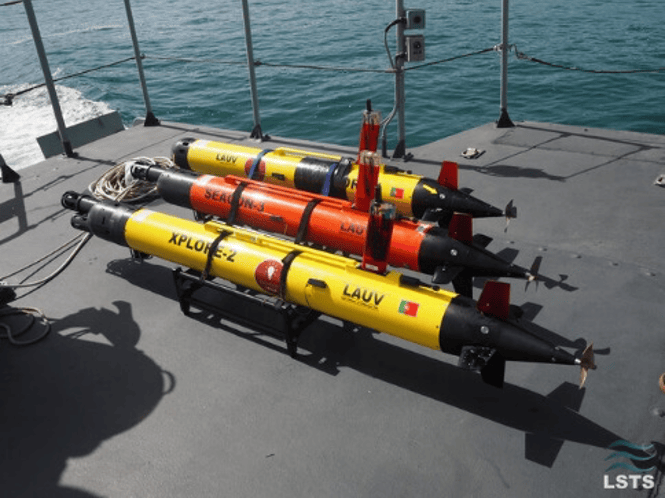}
\caption{Light autonomous underwater vehicles.}
\label{fig:AUV}
\end{figure}
\begin{figure*}[t!]
\centering
\includegraphics[width=0.9\textwidth, trim ={0cm 18.3cm 0cm 3cm}, clip]{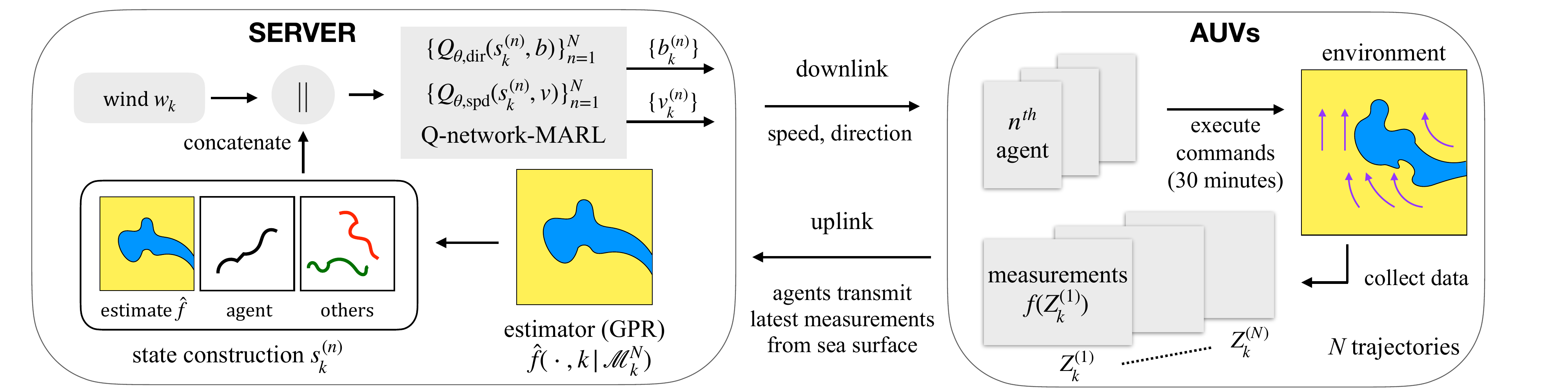} % full page width
\caption{System architecture.}
\label{fig:sys_arc2}
\end{figure*}
A second relevant challenge is that the ocean flow (illustrated in Fig.~\ref{fig:flow}) can significantly impact or even nullify the mobility of an AUV. Note that the ocean flow can push at the same speed as the vehicle's nominal speed of $1$m/s (mobility model in~\eqref{eq:mobility_model}), effectively stalling progress. At the same time, given that the current vector field $c(\cdot)$ is correlated with the salinity field and wind forcing, it may also be exploited for navigation if learned appropriately. 
These two aforementioned challenges strongly motivate the study of a data-driven, long-horizon sequential decision-making solution (e.g., RL), one which we provide in this work. Beyond these constraints, we also address two fundamental engineering aspects. First, multi-agent coordination depends on \textit{communication}, which is intermittent and bandwidth-limited in ocean plume environments. Can we effectively orchestrate multiple AUVs in solving problem~\eqref{eq:setting-obj} with minimal and sporadic communication? Second, endurance hinges on energy efficiency. Can we take sequential decision-making actions that not only aim to keep the mean squared error (MSE) in~\eqref{eq:setting-obj} low, but which also smartly regulate the propulsion used to generate the trajectories $\Gamma_{k}^{(n)}$? Because hydrodynamic drag scales cubically with speed, halving velocity reduces energy consumption by a factor of eight~\cite{carlton2018marine}, thus providing substantial gains in vehicle endurance. In the remainder of the paper, we provide an answer to these questions in the affirmative, illustrating our solution and the results of our simulation study.

\textbf{LAUVs.}
\noindent The sensing agents considered in this work are light autonomous underwater vehicles (LAUVs)~\cite{Teixeria2021}. Each LAUV is equipped with a CTD probe (conductivity, temperature, depth) to measure salinity, navigation sensors, and communication modems (Wi-Fi, GSM, satellite) available when surfacing.
The nominal survey speed of the LAUVs is $1.0$ m/s, yielding an operational endurance of approximately 72 hours. As mentioned in the previous sections, smartly regulating the speed can provide dramatic energy savings, boosting the fleet's endurance. To do this, we introduce a second cruise speed, which we set to $0.4$m/s. Traveling at this speed, the vehicle's endurance can be approximately 8 times longer.
%(potentially bringing endurance to 24 days). 
However, note that the rapid variability of plume dynamics and the strength of coastal currents prevent exclusive reliance on the secondary speed, highlighting the need for adaptive velocity control. 
On the computational side, each vehicle runs lightweight processors sufficient for low-level navigation, sensing and radio communication. We show a picture of LAUVs in Fig.~\ref{fig:AUV}.
\section{Plume Monitoring with MARL}
In this section, we describe the system architecture and the algorithm we design to solve problem~\eqref{eq:setting-obj}. In Fig. \ref{fig:sys_arc2} we provide a graphical illustration of the system architecture, and Algorithm~\ref{alg:plume-mapping} outlines the main control pipeline. 
The key rationales behind the system architecture are (i) offloading computation as much as possible from the AUVs to the server, prioritizing their endurance and (ii) relying only on intermittent and extremely low overhead communication. The pipeline at a given time slot $k$ is as follows: first, all agents dwell on the sea surface and establish a connection with the cloud server via Wi-Fi or GSM. During this communication, each AUV $n=1, ...,N$ uplinks its latest measurements $\{Z_k^{(n)}, f(Z_k^{(n)})\}$ collected over the past 30 minutes. In our implementation, the number of per-agent per-slot measurements $z_{k}^n$ depends on speed and ocean currents, averaging $5$ and limited to a maximum of $10$. Note that transmitting $5$ space-time coordinates and the corresponding measurements only involves transmitting $20$ float numbers, which - in the worst case - only requires $160$ bytes (always $<500$ bytes even with wireless transmission overhead). Upon receiving the new measurements $\{Z_k^{(n)}, f(Z_k^{(n)})\}_{n=1}^N$, the server aggregates them with the previous set, resulting in $\mathcal{M}_k^N = \{D_k^N, f_k^N\}= \mathcal{M}_{k-1}^N\cup \{Z_k^{(n)}, f(Z_k^{(n)})\}_{n=1}^N$ and updates the salinity map estimate $\hat{f}(\cdot, \sigma[k]|\mathcal{M}_k^N)$. Based on $\hat{f}$, the current agent trajectories $D_k^N$, and a vector containing the wind speed and direction $w_k$, the server determines the control policy $\pi(\hat{f}, \mathcal{M}_k^N, w_k)$ for all AUVs $n=1, ..., N$ for the next 30 minutes time slot, which results in direction $\{b_k^{(n)}\}_{n=1}^N$ and speed levels $\{v_k^{(n)}\}_{n=1}^N$. At that point, the server downlinks the commands to the AUVs, which resubmerge and resume sampling, starting to build $\{Z_{k+1}^{(n)}, f(Z_{k+1}^{(n)})\}_{n=1}^N$. The estimator $\hat{f}(\cdot, \sigma[k]|\mathcal{M}_k^N)$ and the policy $\pi(\hat{f}, \mathcal{M}_k^N, w_k)$ form the core of our framework. We detail these two components in the following subsections, respectively.\vspace{-0.1cm}
\begin{algorithm}[t]
\caption{Plume-DQN-GP}
\label{alg:plume-mapping}
\begin{algorithmic}[1]
\STATE \textbf{Input:} %window $M$, 
%horizon $T$, 
AUVs $n=1, ..., N$, action space ($\mathcal{H}$, $\mathcal{V}$), 
%step $\Delta$, 
hyperparameters
\STATE \textbf{Initialization:} deploy $N$ AUVs in predetermined initial locations, directions $b_0^n$ and speeds $v_0^n$ for $n = 1, ..., N$ 
%build initial GP map $\hat f_0$ from $(D_{0;M},f_{0;M})$
    \FOR{$k = 1,\ldots,T$}
        \FOR{$n = 1,\ldots,N$ \text{in parallel}}
        \STATE The $n$-th agent receives $(b_{k-1}^{(n)}, v_{k-1}^{(n)})$ commands from server and executes, visiting trajectory locations $Z_k^{(n)}$ and collecting measurements $f(Z_k^{(n)})$ for 30 minutes, and then transmits sets $\{Z_k^{(n)}, f(Z_k^{(n)})\}$ to server from sea surface. 
        \ENDFOR
        \STATE Server receives $\{Z_k^{(n)}, f(Z_k^{(n)})\}_{n=1}^N$, aggregates $\mathcal{M}_k^N = \{D_k^N, f_k^N\}= \mathcal{M}_{k-1}^N\cup \{Z_k^{(n)}, f(Z_k^{(n)})\}_{n=1}^N$, and uses it to update $\hat{f}(\cdot, \sigma[k]|\mathcal{M}_k^N)$.
        \STATE Server constructs per-agent state embeddings $s_k^{(n)}$ from $\hat f (\cdot, \sigma[k]|\mathcal{M}_k)$, trajectories $D_k$ and wind vector.
        \STATE Server computes directions and speeds $\{b_k^{(n)}, v_k^{(n)}\}$  for $n =1, ..., N$ using the Q-network, see~\eqref{eq:Q_policy}.
%        \[
%        b^n_t \in \arg\max_{b\in\mathcal{H}} Q_{\mathrm{dir}}(s_t^n,b),
%        \quad
%        v^n_t \in \arg\max_{v\in\mathcal{V}} Q_{\mathrm{spd}}(s_t^n,v)
%        \]
        \STATE Server transmits $\{b_k^{(n)}, v_k^{(n)}\}$ in downlink to agents. 
    \ENDFOR
\end{algorithmic}
\end{algorithm}
\subsection{Estimation Module}\label{sec:estimation_module}

To produce measurement-based estimates $\hat{f}(\cdot, \sigma[k]|\mathcal{M}_k^N)$ we resort to non-parametric Gaussian process regression (GPR). GPR represents a very appealing choice in this case because (i) it encodes spatial and temporal correlations through a simple, interpretable kernel function; (ii) it is highly flexible and reliable when in need of interpolating the sparse and highly irregular samples $\mathcal{M}_k^N$ produced by the trajectory-constrained AUVs. Our GPR estimator places a spatiotemporal Gaussian prior on the salinity field of interest
\begin{equation}
f(x,t)\;\sim\;\mathcal{GP}\!\big(m(x,t),\,K\big((x,t),(x',t')\big)\big), 
\end{equation}
\noindent with mean function $m(\cdot)$ and covariance kernel $K(\cdot, \cdot)$. In this work, we fix the mean $m(x, t) = f_{\text{ocn}}$. Given  the set  $\mathcal{M}_k^N$, the posterior mean at $(x,t)$ can be computed in closed form as:
\begin{equation}
\begin{aligned}
\hat{f}(x,t \mid \mathcal{M}_k^N) 
= f_{\text{ocn}} + k_\ast \bar{K}^{-1}\!\big(f(D_k^N) - f_{\text{ocn}}\big)
\end{aligned}
\label{mean}
\end{equation}
where $k_\ast=K\!\big((x,t),D_k^{(N)}\big)$,  %\in\mathbb{R}^{1\times |D_k^{(N)}|}
$\bar{K}=K\!\big(D_k^{(N)},D_k^{(N)}\big)+\sigma^2 I $ are the $(x,t)$-kernel section and the kernel data matrix, respectively, with $\sigma^2$ capturing the measurement noise. To keep computations tractable over long horizons, we retain only the most recent $M$ slots in $\mathcal{M}_k^N$, discarding older samples.

\begin{figure}
\centering
\includegraphics[width=0.8\columnwidth, trim ={0.3cm 0.3cm 0.2cm 0.2cm}, clip]{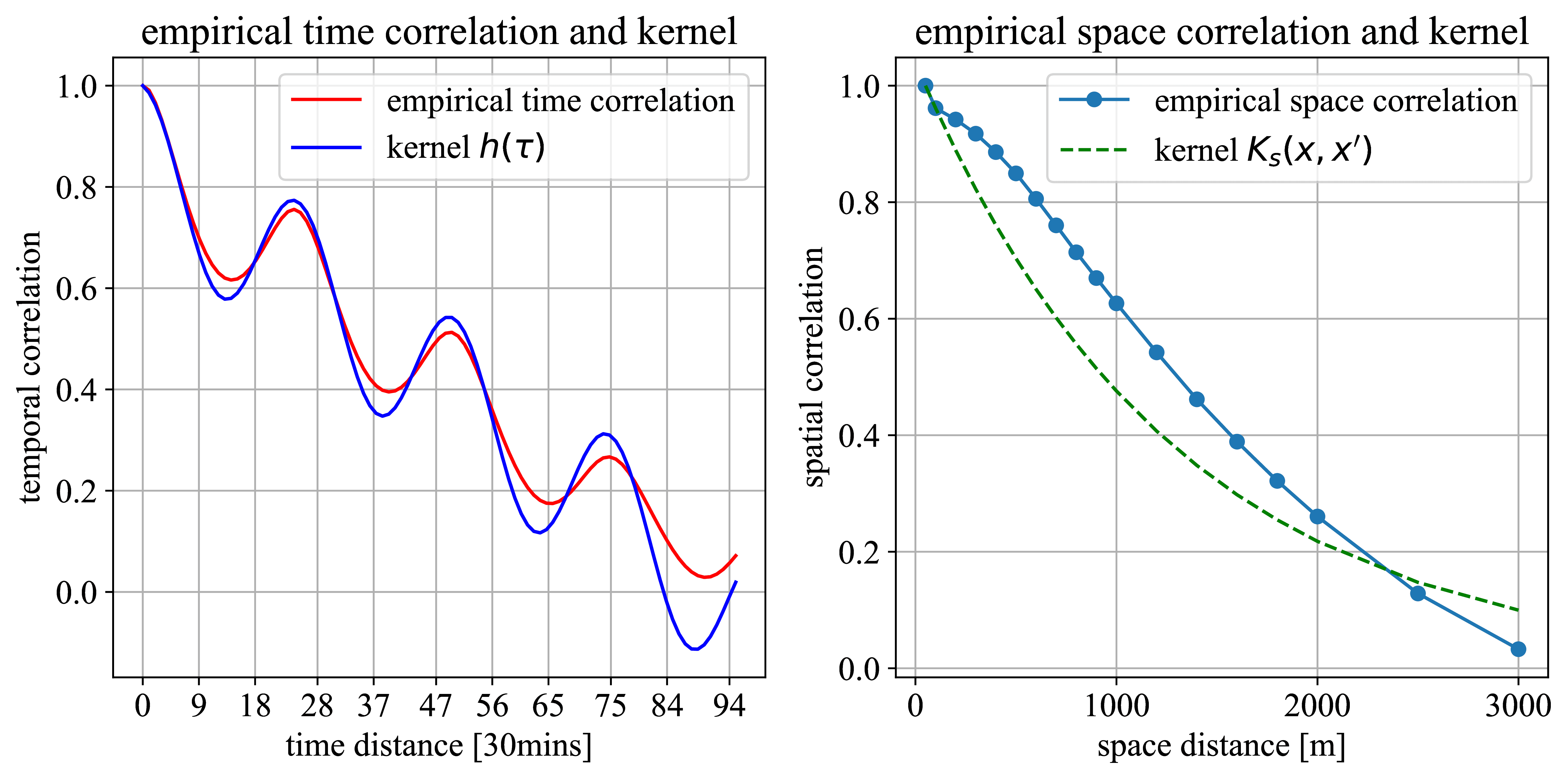}
	\caption{{Temporal and spatial kernels $h(\tau)$ and $K_s(x, x')$ fitting the corresponding empirical correlations.}}
 \label{fig:kernel}
\end{figure}

\textbf{Spatiotemporal kernel.}
We adopt a space-time separable spatiotemporal kernel of the following form:
\begin{equation}
K\big((x,t),(x',t')\big)=K_s(x,x')\,h\left(\tau\right),
\end{equation}
where $\tau = |t-t'|$. We choose the functions $K_s(x,x')$ and $h\left(|t-t'|\right)$ by analyzing and fitting the empirical correlations from the historical data. In particular, we set
\begin{equation}
K_s(x,x')=\lambda^2\exp\!\left(-\frac{\|x-x'\|}{\,\ell}\right),
\end{equation}
with hyperparameters $\lambda, \ell >0$ and 
\begin{equation}
\begin{aligned}
 h(\tau)=\beta_0-\beta_1\,\tau+\beta_2\Big(\cos\!\big(\pi\tau/T_0\big)-1\Big),  
 \end{aligned}
\end{equation}
which combines a linearly decaying term and a periodic oscillation, whose period $T_0$, not surprisingly, is the same as the tidal period of 12.5 hours. 
All hyperparameters are fit to historical data by matching empirical correlations, as we illustrate in Figure~\ref{fig:kernel}. In the implementation, we made sure that the hyperparameters and memory window $M$ ensure positive definiteness of the kernel.

\medskip

\subsection{Decision Making Module}
We cast the construction of trajectories $D_k^N$ over time slots $k=1,..., T$ to solve~\eqref{eq:setting-obj} as a (centralized) multi-agent sequential decision-making problem. In particular, we want to design a decision making policy $\pi(\cdot)$ that, at each time slot $k$, given measurements and trajectory history in $\mathcal{M}_k^N$, and exogenous inputs (e.g., wind), selects a direction and a speed level for each AUV for the next time slot (30 minutes). Given the large amount of historical environmental data and the availability of the advanced Delft3D simulator, a natural choice to learn $\pi(\cdot)$ is model-free multi-agent reinforcement learning (MARL)~\cite{sutton1998reinforcement}. Due to the well-known problem of \textit{centralized} MARL lacking scalability\cite{li2020deep}, we adopt the typical solution of treating the problem as if we were in a decentralized MARL setting with full communication~\cite{vinyals2019grandmaster}. To do so, we need to define a per-agent state space $\mathcal{S}$ and action space $\mathcal{A}$, and a reward function $R:\mathcal{S}\times\mathcal{A}\rightarrow\mathbb{R}$.
We seek a policy map $\pi: \mathcal{S} \rightarrow \mathcal{A}$ that maximizes the expected cumulative discounted reward $\mathbb{E}\left[\sum_{k = 0}^\infty \gamma^k R(s_k, a_k)\right]$. At each time step $k$, the server builds the state $s_k^{(n)}$ for each agent using the current estimate $\hat{f}$ and the set $D_k$, and then selects an action $a_k^{(n)}$, for $n=1, ..., N$. To learn policy $\pi(\cdot)$, we use Q-learning \cite{sutton1998reinforcement},
which, for a given discount factor $\gamma>0$, 
targets the optimal action–value function
\begin{equation}
Q^{\star}(s,a) = \mathbb{E}\Bigl[\, r + \gamma \max_{a'} Q^{\star}(s',a') \,\big|\, s,a \Bigr],    
\end{equation}
and uses it to select actions. Since the state space is high-dimensional, we approximate $Q(s,a)$ with a neural network with parameters $\theta$ and denote it by $Q_{\theta}(s,a)$.

We factor the action space $\mathcal{A}$ into discrete direction and speed components
\begin{equation}
\begin{aligned}
a &= (b,v) \;\in\; \mathcal{H}\times\mathcal{V} = \mathcal{A},\ \text{where}\\
\mathcal{H}&=\{0,45, ..., 315\}\ \deg, \ \mathcal{V}=\{0.4,\,1.0\}\ \text{m/s}. 
\end{aligned}
\end{equation}
The two velocity levels enable us to control the trade-off between endurance and agility: the lower cruise speed ($0.4$ m/s) extends mission duration by conserving propulsion energy, while the higher speed ($1.0$ m/s) enables more accurate plume monitoring at the expense of battery life.
%This explicit modeling of speed in the action space allows the policy to learn when to prioritize coverage versus conservation.

To improve scalability, we decouple the choice of speed and direction, decomposing $Q_\theta$ into two components, each sharing a dueling architecture~\cite{wang2016dueling} with state-value function $V_{\bar{\theta}, \theta_1}(s)$ and advantage functions $A^{\text{dir}}_{\bar\theta,\theta_2}(s,b)$: $A^{\text{spd}}_{\bar\theta,\theta_3}(s,v)$,
\begin{equation*}\label{eq:Qangle}
\begin{aligned}
Q_{\theta,\mathrm{dir}}(s,b) &= V_{\bar{\theta}, \theta_1}(s) + A^{\text{dir}}_{\bar\theta,\theta_2}(s,b) - \frac{1}{|\mathcal{H}|}\sum_{b'\in\mathcal{H}} A^{\text{dir}}_{\bar\theta,\theta_2}(s,b')    \\
Q_{\theta,\mathrm{spd}}(s,v) &= V_{\bar\theta, \theta_1}(s) + A^{\text{spd}}_{\bar\theta,\theta_3}(s,v) - \frac{1}{|\mathcal{V}|}\sum_{v'\in\mathcal{V}} A^{\text{spd}}_{\bar\theta,\theta_3}(s,v')
\end{aligned},
\end{equation*}
where note that $\theta = \{\bar{\theta}, \theta_1, \theta_2, \theta_3\}$.
For agent $n\in\{1, ..., N\}$ in state $s_k^{(n)}$, action selection is then performed per head:
\begin{equation}\label{eq:Q_policy}
\begin{aligned}
    b^{(n)}_k &= \arg\max_{b\in\mathcal{H}} Q_{\theta,\mathrm{dir}}(s_k^{(n)},b),\\
v^{(n)}_k &= \arg\max_{v\in\mathcal{V}} Q_{\theta,\mathrm{spd}}(s_k^{(n)},v),
\end{aligned}
\end{equation}
which establishes direction and speed.

\textbf{State construction.} 
The state representation $s_k^{(n)}$ comprises three components. Following prior work on image-based policy inputs \cite{ Kim2023}, we render both spatial estimates and trajectory traces into a fixed-resolution $3$-channels image, compressed by a convolutional neural network (CNN) into a compact embedding. Specifically, the three channels for the $n$-th agent are
%• \textbf{current map estimate channel}: 
(i) a compressed image of the GP estimate of the salinity field $\hat{f}(\cdot,\sigma[k])$, 
%• \textbf{trajectory imprints}: 
(ii) one image marking in white over a black background recent measurement locations of the agent, $Z_k^{(n)}, Z_{k-1}^{(n)}, ...$ (iii) one image marking in white over a black background recent measurement locations for agent $n$ teammates, $\{Z_k^{(l)}, Z_{k-1}^{(l)}, ...\}_{l\neq n}$. The trajectory marks intensities are log-scale weighted by recency.
In addition, a
%• \textbf{wind forcing}: 
wind vector $w_k = [\text{angle}, \text{speed}]$, processed by a NN to produce an embedding, is concatenated with the rest of the network. See Fig.~\ref{fig:sys_arc2} for an illustration of the state construction. Including the GP variance in the state did not provide improvements in our experiments, so we do not include it.

\textbf{Reward design.} We design a reward function which combines cooperative and individual components. This incentivizes agents to {cooperate}  by minimizing the global MSE while also seeking individual credit for reducing MSE along their own trajectory. Indeed, experimenting with a purely cooperative global reward, we incurred in the well-known \textit{credit assignment} problem~\cite{nguyen2018credit}. Let us introduce a salinity contrast score
\begin{equation}
    F_k \;=\; \Bigl|\,f_{\mathrm{ocn}}-\tfrac{1}{|G|}\sum_{x\in G} f(x,\sigma[k])\Bigr|, \quad f_{\mathrm{ocn}}\approx 35~\text{psu},
\end{equation}
which characterizes how much freshwater is present in the ocean at time slot $k$.
\begin{figure}[t]
\centering
\includegraphics[width=0.8\columnwidth, trim ={0cm 0.0cm 0cm 0cm}, clip]{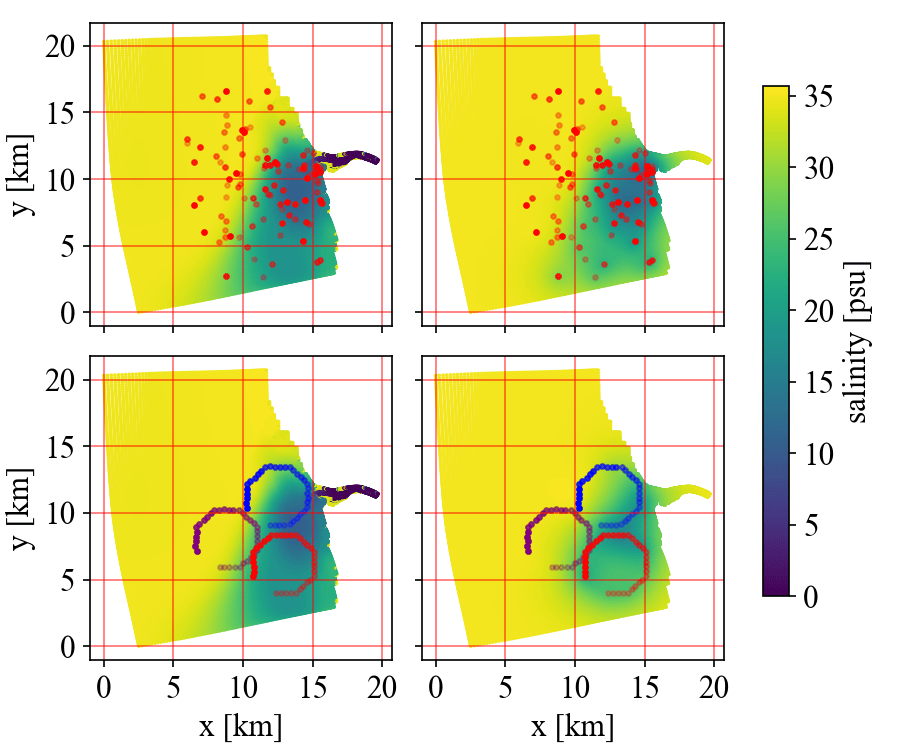} %PLACEHOLDER
\caption{Left column and right column: ground-truth and estimated salinity field, respectively. On top: uniform random sampling on the non-homogeneous spatial grid, with measurement locations denoted as red points. At the bottom, example of trajectory-constrained measurements: fixed rotations' trajectories of three agents (in different colors).}
\label{fig:unif_rots}
\end{figure}
Our reward function for agent $n$ is
\begin{equation}\label{eq:reward}
\begin{aligned}
r_k^{(n)} &= r_{k,g} + r_{k,n}, \ \text{where}
\\
r_{k,g} &= -\eta_0e_k^G + \eta_1\frac{F_k}{1+e_k^G} - \eta_2\sum_{n=1}^Nv^{(n)}_k,
\\
r_{k,n} &= \eta_3\sum_{(x, t) \in Z_k^{(n)}} \left( \hat{f}_{k-1}(x) - f(x, \sigma[k]) \right)^2,
\end{aligned}
\end{equation}
where $e_k^G$ is the global MSE in the evaluation grid $G$, $\hat{f}_{k-1}(x) = \hat{f}(x, \sigma[k-1]|\mathcal{M}_{k-1}^N)$ and $\eta_0, ..., \eta_3>0$ are hyperparameters.  The term $r_{k,n}$ is the key individual credit that the agent earns for visiting locations $x$ along its trajectory $Z_k^{(n)}$, where the previous estimate $\hat{f}_{k-1}(x)$ was significantly inaccurate relative to the actual field $f(x, \sigma[k])$. The term $\sum_{n=1}^Nv^{(n)}_k$ penalizes higher speed at a fleet level. The salinity contrast score $F_k$ incentivizes a reduction in the MSE when there is more freshwater in the plume. 
\section{Simulation Study and Results}
\begin{figure}[t]
\centering
\includegraphics[width=0.9\columnwidth, trim ={0.26cm 0.0cm 0.2cm 0.2cm}, clip]{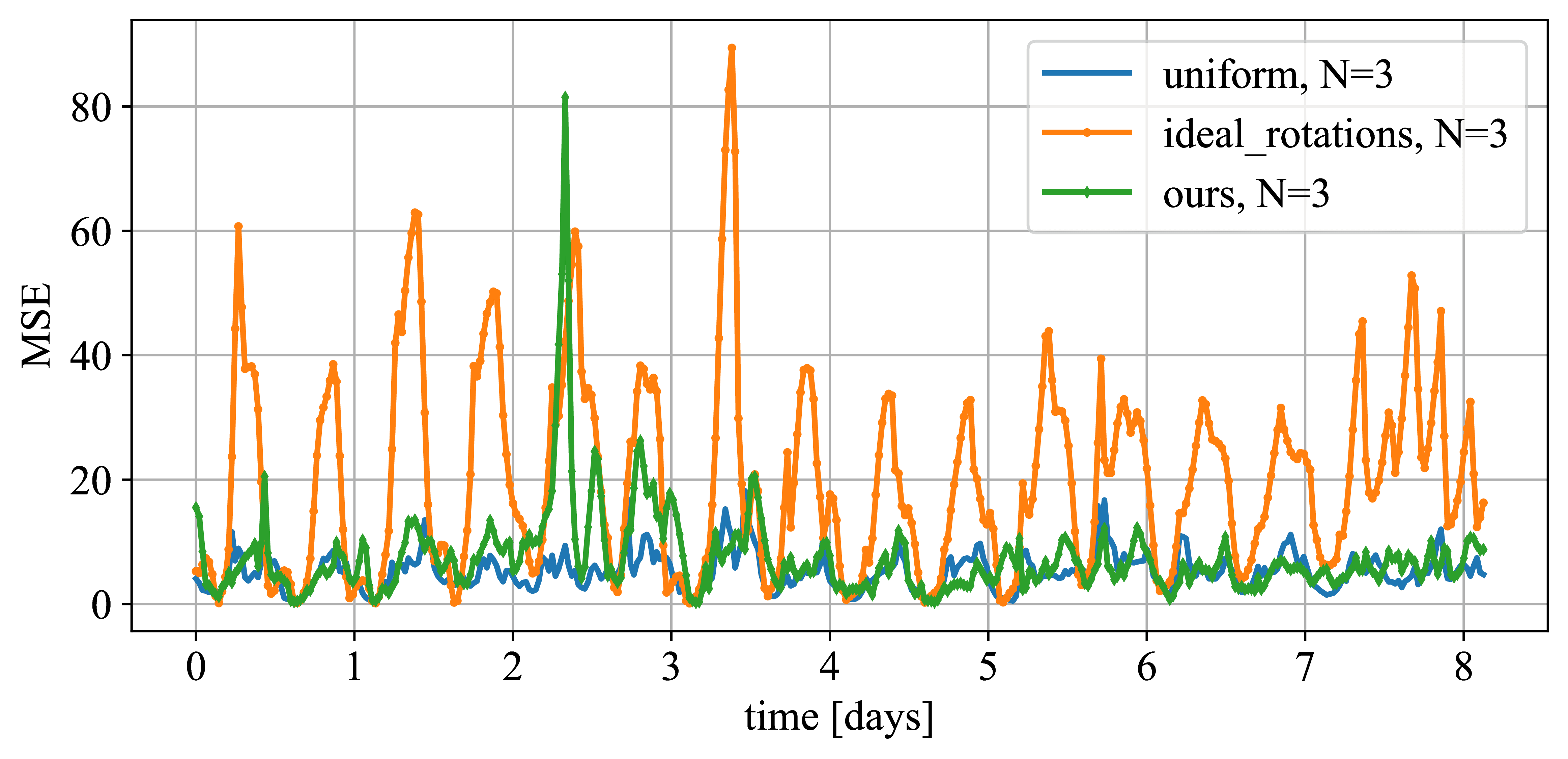} %PLACEHOLDER
\caption{Comparison of MSE over time for the two baselines (uniform and rotations) and our algorithm, interval of 8 days beginning of February 2018}
\label{fig:MSE_unif_rots}
\end{figure}
In this section, we illustrate the details of our simulation study and of our numerical results.

\textbf{Dataset.}  We train our algorithm over a selection of $4$ months data from the year 2018 (February, April, October and December) and evaluate over different months from 2016-2018 (see Table~\ref{tab:season-vertical}). Overall, we use a total of $\approx 6k$ time frames for the training, with a $30$ minutes time resolution. Each frame spans $\approx5\times10^4$ grid locations and measurement noise is i.i.d.\ $\mathcal{N}(0,\,0.01)$.

\textbf{Training setup. } We minimize the Bellman loss for the Q-network, computed over mini-batches of size $64$ sampled from a replay buffer. We use Adam optimizer, and we perform soft updates using a target network. Each policy used in this section has been trained for $6500$ episodes of length $150$ ($\approx 3$ days) and a discount factor $\gamma = 0.9$, with the first $1500$ episodes of pure exploration. The Q-network uses a 3-layer CNN followed by two fully connected layers.

\textbf{Two baselines.} 
In Fig.~\ref{fig:unif_rots}, we illustrate two different sampling schemes on ground-truth salinity fields $f$ (left) and their GP estimates $\hat f$ (right) for October 2018. Both methods use the same budget of 15 measurements every 30 minutes with a memory window of 24 frames, and all estimates employ the GPR model of Sec.~\ref{sec:estimation_module}.  The top panels show an unconstrained uniform strategy: measurements are placed uniformly at random in the spatial non-homogeneous grid of the Delft3D numerical model, which has a higher density of grid points in higher variance areas. In practice, this effectively biases sampling toward high-variance locations closer to the river mouth. Note that since uniform placement ignores the trajectory constraint $Z_k^{(n)}\subset\Gamma_{k}^{(n)},\ \forall k,n$, it is not physically realizable with the AUVs, but does provide a fundamental baseline. If one could freely deploy measurements without vehicle constraints, this uniform scheme would represent a natural choice for estimating $f$. In the bottom panels of Figure~\ref{fig:unif_rots}, we show an example of a predetermined strategy which is compliant with the constraints in~\eqref{eq:setting-obj}. This particular strategy is based on rotations around some strategic core points in the plume, and, for the sake of this illustration, we assume that the robots are able to perform the rotations without being impacted by the ocean currents. Although the trajectories are strategically placed in key areas, one can already see from Fig.~\ref{fig:unif_rots} how the constrained sampling estimate struggle to capture the state of the plume, while the uniform sampling strategy provides a better estimation. To illustrate this numerically, in Figure~\ref{fig:MSE_unif_rots} we show the evolution of the MSE over time for these two baselines (``uniform'' and ``ideal rotations''). One can see the notable difference in terms of MSE for the two different strategies: the MSE of the rotations is roughly 6 times higher than the unconstrained uniform baseline under the same sampling budget.
\begin{figure}[t]
\centering
\includegraphics[width=0.9\columnwidth, trim ={0cm 0.2cm 0cm 0.1cm}, clip]{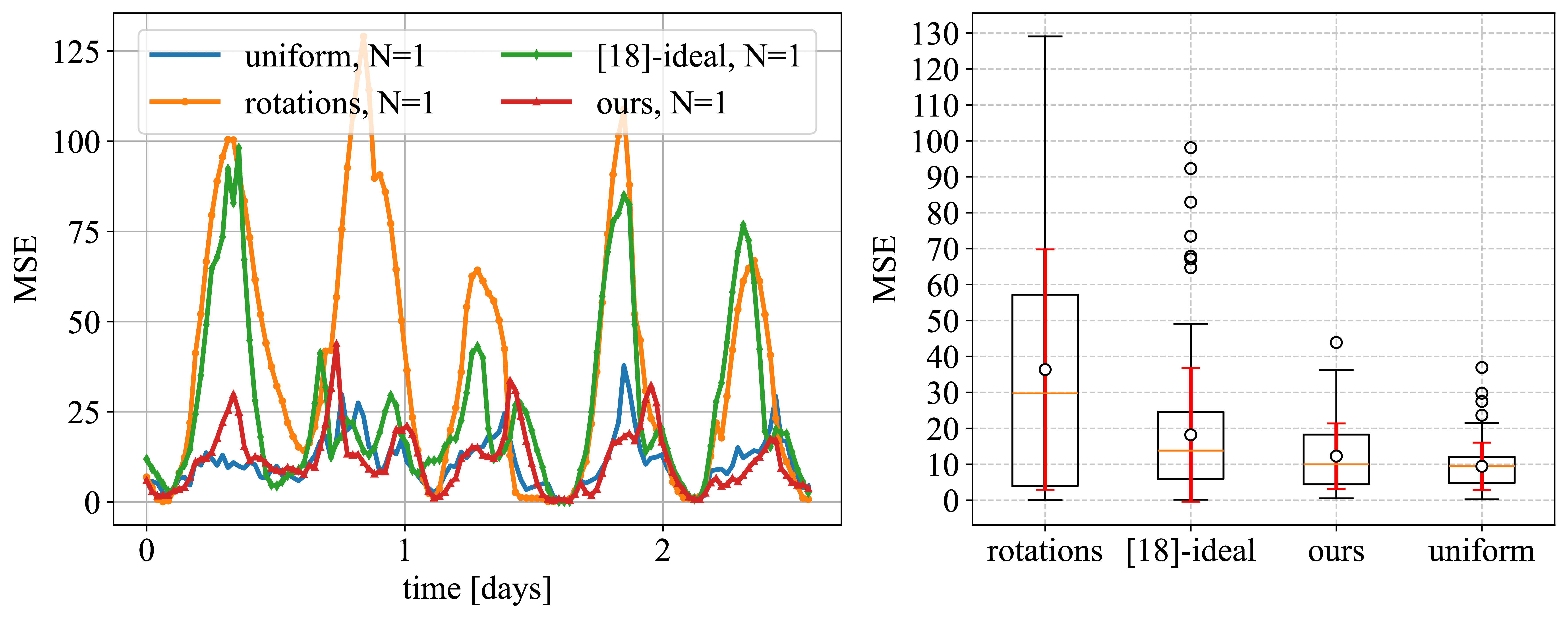} %PLACEHOLDER
\caption{Example of mapping performance when using a \textit{single} ($N=1$) AUV. Comparison between our algorithm, baselines and the benchmark~\cite{Berild2024}. On the left, an example of the MSE evolution over time for a small time interval ($\approx$ 3 days). On the right, box plots of the MSE over time obtained by the algorithms executed over the whole month of February 2018.}
\label{fig:single_agent_comp}
\end{figure}
\begin{figure}[t]
\centering
\includegraphics[width=0.95\columnwidth, trim ={0cm 0.2cm 0cm 0.1cm}, clip]{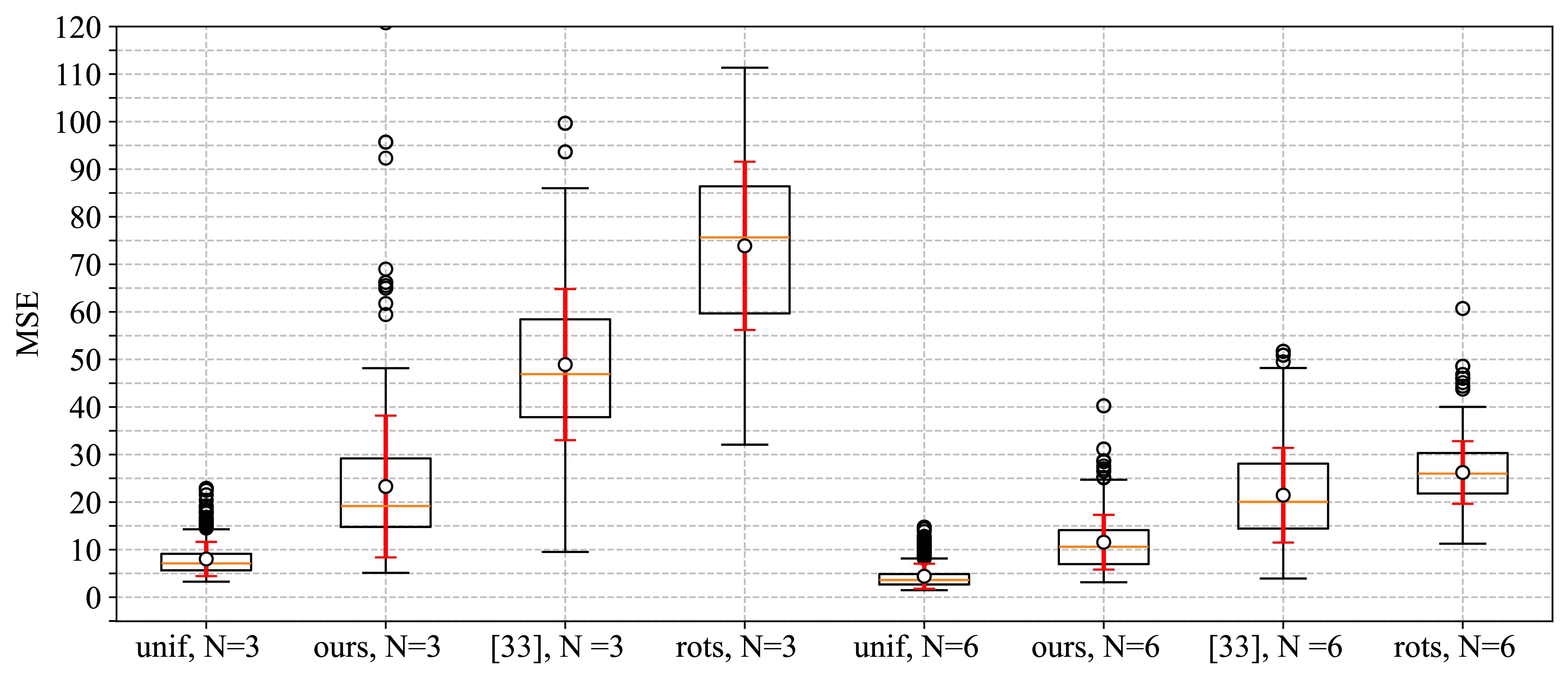}
\caption{Multi-agent performance for the test month of March 2018. ``unif" and ``rots" stand for uniform and rotations baselines, respectively.}
\label{fig:MAmse_comp}
\end{figure}

\textbf{Single-agent performance.}
First, we analyze the performance of our algorithm when deploying only one AUV $N=1$ to map the river plume. This allows us to compare our solution directly with the one proposed in~\cite{Berild2024}, which only considered a single-agent setting, and which
is based on evaluating the expected integrated Bernoulli variance (EIBV) for each potential direction at every step. For this benchmark, we implement an \textit{ideal} version in which the decision maker has access to the actual IBV for each candidate direction, and we term this algorithm ``ideal~\cite{Berild2024}". This allows us to compare against the best possible version of~\cite{Berild2024} and to obtain results which do not depend on the specific numerical way of computing the EIBV. A fundamental parameter of~\cite{Berild2024} is a salinity threshold that defines the  plume front. While the original paper fixed this at 24 psu, we found this too low for the Douro plume and selected the best threshold from $\{28,30,32,34\}$, choosing 32 psu. We show an example of simulation outcome in Figure~\ref{fig:single_agent_comp}, where on the left, we plot MSE evolution over three days for four algorithms, and on the right, we report monthly MSE box plots for February 2018. The two plots reveal how, despite~\cite{Berild2024} performing well at certain parts of the process, it tends to lose track of the plume over the long run, while our RL policy, in contrast, consistently maintains low MSE. This result is not surprising, as the quick temporal evolution of the plume can make the decisions taken by~\cite{Berild2024} based on the current prior Gaussian field myopic with respect to the spatiotemporal evolution of the plume,  whereas our learned policy explicitly optimizes long-horizon mapping performance.
\begin{table}[t]
\centering
\caption{Policy generalization across   seasonal regimes}
\label{tab:season-vertical}
\small
\setlength{\tabcolsep}{4pt}
\renewcommand{\arraystretch}{1.15}
\resizebox{\columnwidth}{!}{
\begin{tabular}{@{}l
S[table-format=1.2] S[table-format=1.1] S[table-format=1.2]
S[table-format=1.2] S[table-format=1.1] S[table-format=1.2]@{}}
\toprule
& \multicolumn{3}{c}{\textbf{$N=3$}} & \multicolumn{3}{c}{\textbf{$N=6$}} \\
\cmidrule(lr){2-4}\cmidrule(lr){5-7}
Month & {\textbf{Ours} ${\mathrm{MSE}}$} & {End. (d)} & {\cite{pratissoli2025distributed} ${\mathrm{MSE}}$}
      & {\textbf{Ours} ${\mathrm{MSE}}$} & {End. (d)} & {\cite{pratissoli2025distributed} ${\mathrm{MSE}}$} \\
\midrule
Mar ’18 & \bfseries 23.14 & 3.5 & 47.87 & \bfseries 11.46 & 3.1 & 21.07 \\
Sep ’18 & \bfseries 10.36 & 4.6 & 18.48 & \bfseries 5.44 & 5.0 & 6.27 \\
Nov ’18 & \bfseries 13.96 & 4.2 & 20.99 & \bfseries 7.88 & 4.3 & 8.24 \\
Jan ’16 & \bfseries 27.03 & 3.2 & 49.19 & \bfseries 12.84 & 3.3 & 19.53 \\
Feb ’16 & \bfseries 24.26 & 3.3 & 49.74 & \bfseries 14.07 & 3.4 & 21.32 \\
Oct ’17 & \bfseries 3.69 & 13.0 & 4.04 & \bfseries 2.79 & 15.6 & 2.04 \\
\bottomrule
\end{tabular}}
\end{table}
\begin{figure}[t]
\centering
\includegraphics[width=0.7\columnwidth, trim ={.4cm 0.3cm .4cm 0.3cm}, clip]{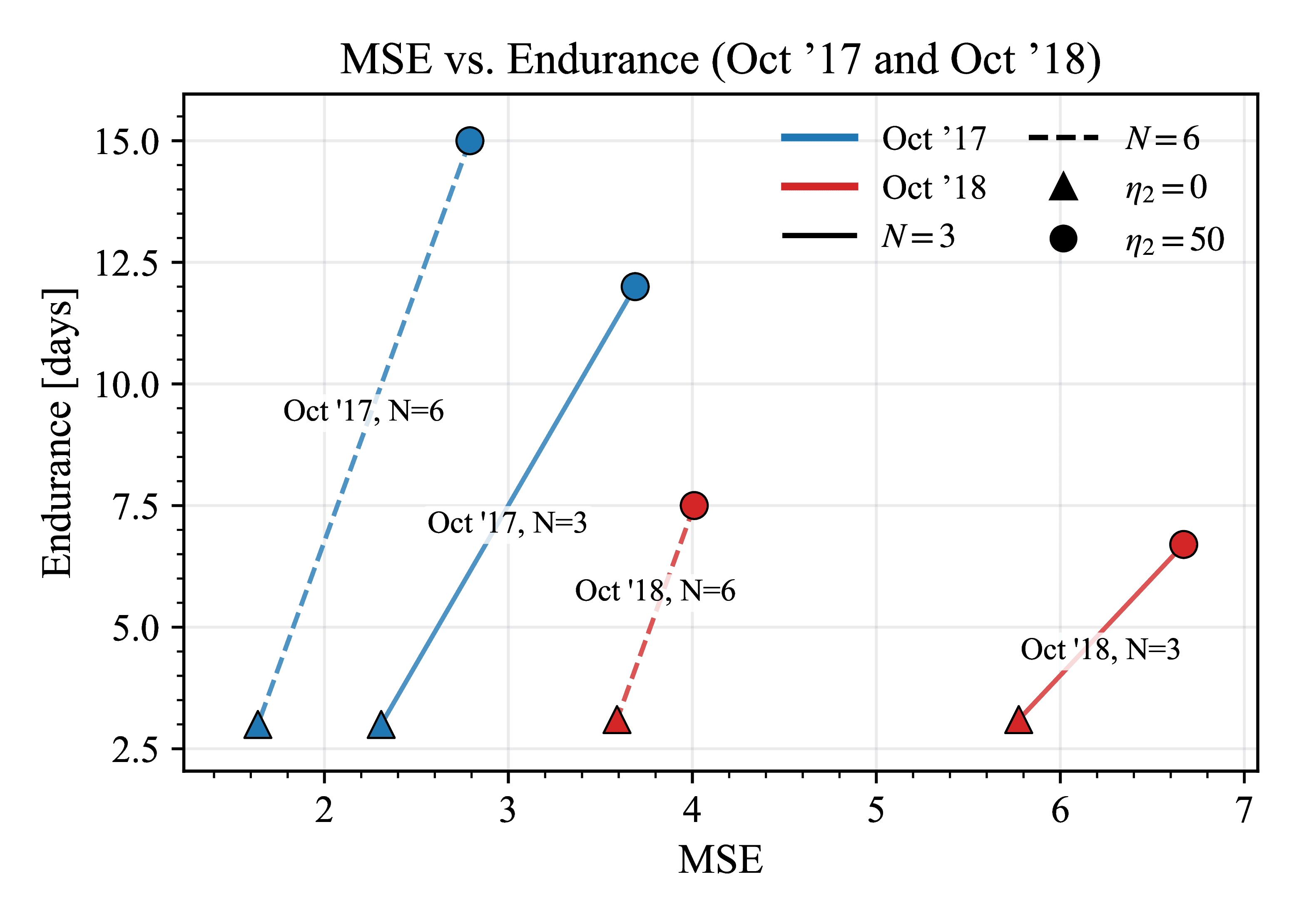}
\caption{Endurance-MSE tradeoff for the representative month of October.}
\label{fig:tradeoff}
\end{figure}
\begin{figure*}[t]
\centering
\includegraphics[width=0.95\textwidth, trim ={7cm 3.5cm 7cm 3cm}, clip]{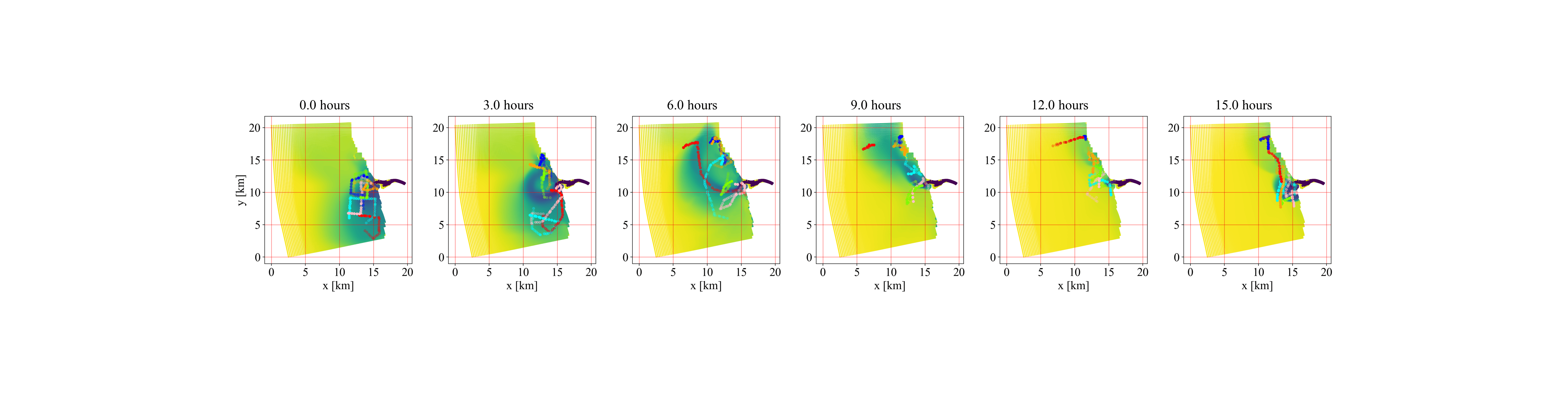} % full page width
\caption{Example of the multi-agent control performance of our solution with $N=6$ agents over a 15 hours time window, in the testing month of March 2018. The colored points show the different agents' trajectories and corresponding measurements' locations.}
\label{fig:control_frames_seq}
\end{figure*}

\textbf{Multi-agent performance.}
We now evaluate performance in the multi-agent setting, focusing on: (i) MSE gains provided by the multi-agent coordination when increasing $N$, and on the comparison with existing literature benchmarks; (ii) the energy efficiency improvements induced by varying the $\eta_2$ reward hyperparameter in~\eqref{eq:reward} and the corresponding MSE-energy tradeoff. As a multi-agent benchmark, we implement the recent work in~\cite{pratissoli2025distributed} that proposed an algorithm based on adaptive Voronoi partitioning of the space of interest to track a time-varying process. We empirically tune the exploration-exploitation hyperparameters in~\cite{pratissoli2025distributed} and show the best results in terms of MSE. In Fig.~\ref{fig:MAmse_comp}, we compare the different schemes for $N=3$ and $N=6$ for March 2018. We first note how the adaptive Voronoi partition solution~\cite{pratissoli2025distributed} significantly outperforms the strategic rotations solution. We also note that our algorithm provides maps with half of the MSE compared to~\cite{pratissoli2025distributed}, for both $N=3$ and $N=6$. In Fig.~\ref{fig:control_frames_seq}, we show a sequence of frames to illustrate the multi-agent control performance of the agents with respect to the salinity field $f$ evolution over a $15$ hour time interval. In Fig.~\ref{fig:tradeoff} we show the endurance–MSE trade-off obtained by varying the reward weight $\eta_2$ in~\eqref{eq:reward}, using October 2017 and October 2018 as exemplars. Increasing $\eta_2$ biases the policy toward lower propulsion use, thus extending endurance at a modest cost in accuracy, while scaling from $N=3$ to $N=6$ both reduces MSE and more than doubles endurance. Table~\ref{tab:season-vertical} summarizes results obtained with $\eta_2 = 50$, and further confirms that our algorithm generalizes across unseen (during training) years and seasonal regimes.

\section{Conclusion and Future Work}
We conducted a simulation study showing promising performance in mapping the Douro river plume with multiple underwater vehicles over long time horizons. Future works include considering 3D mapping and real-world deployment of the LAUVs robots in the river plume.
\bibliographystyle{IEEEtran}
\bibliography{ref}
\end{document}